\documentclass[
twocolumn,
]{ceurart}

\usepackage{listings,multirow,multicol}
\begin{document}

\copyrightyear{2022}
\copyrightclause{Copyright for this paper by its authors.
  Use permitted under Creative Commons License Attribution 4.0
  International (CC BY 4.0).}

\conference{IJCAI 2023 Workshop on Deepfake Audio Detection and Analysis (DADA 2023), August 19, 2023, Macao, S.A.R}

\title{Low-rank Adaptation Method for Wav2vec2-based Fake Audio Detection}


\author[1,2]{Chenglong Wang}[%
email=chenglong.wang@nlpr.ia.ac.cn,
]
\address[1]{University of Science and Technology of China, Hefei, China}
\address[2]{State Key Laboratory of Multimodal Artificial Intelligence Systems, Institute of Automation, Chinese Academy of Sciences, Beijng, China}

\author[2,3]{Jiangyan Yi}[%
email=jiangyan.yi@nlpr.ia.ac.cn,
]
\cormark[1]
\address[3]{School of Artificial Intelligence, University of Chinese Academy of Sciences, China}

\author[2,4]{Xiaohui Zhang}[%
email=21120320@bjtu.edu.cn,
]
\address[4]{School of Computer and Information Technology, Beijing Jiaotong University}

\author[3,5]{Jianhua Tao}[%
email=jhtao@tsinghua.edu.cn,
]
\cormark[1]
\address[5]{Department of Automation, Tsinghua University}

\author[2]{Le Xu}

\author[2,3]{Ruibo Fu}

\cortext[1]{Corresponding author.}
\fntext[1]{These authors contributed equally.}

\begin{abstract}
Self-supervised speech models are a rapidly developing research topic in fake audio detection. Many pre-trained models can serve as feature extractors, learning richer and higher-level speech features. However,when fine-tuning pre-trained models, there is often a challenge of excessively long training times and high memory consumption, and complete fine-tuning is also very expensive. To alleviate this problem, we apply low-rank adaptation(LoRA) to the wav2vec2 model, freezing the pre-trained model weights and injecting a trainable rank-decomposition matrix into each layer of the transformer architecture, greatly reducing the number of trainable parameters for downstream tasks. Compared with fine-tuning with Adam on the wav2vec2 model containing 317M training parameters, LoRA achieved similar performance by reducing the number of trainable parameters by 198 times.
\end{abstract}


\maketitle

\section{Introduction}
Self-supervised pre-training has become a popular research topic in recent years and has already been applied in the fields of natural language processing and computer vision. In the domain of speech processing, numerous self-supervised speech models have been proposed, such as wav2vec \cite{schneider2019wav2vec}, wav2vec2\cite{baevski2020wav2vec}, and HuBERT \cite{hsu2021hubert}, which enables learning of high-level representations of the speech signal. These models have been successfully applied in various areas, including speech recognition, speaker recognition, and emotion recognition \cite{fan2020exploring, pepino2021emotion}. However, in the realm of fake audio detection, only a limited number of studies have explored the use of pre-trained models as feature extractors. Specifically, some researchers have employed the pre-trained wav2vec2 model, which has demonstrated the ability to extract more robust deep features \cite{xie2021siamese}, and has been applied in the recent ADD2022 challenge \cite{yi2022add, martin2022vicomtech}. This model takes the raw waveform as input and learns speech information from a large amount of unlabeled speech data, potentially containing valuable information for identifying fake audio. In addition to wav2vec2, other studies have investigated various self-supervised models as potential feature extractors for the spoof detection task \cite{wang2021investigating}.

The method of using wav2vec2 as a feature extractor typically requires fine-tuning on the training set. This process requires updating all parameters of the wav2vec2 model to create a new model that contains the same parameters as the original one. This process requires a large amount of computing resources for training as well as specialized graphics memory. For example, the Wav2vec2 XLSR \cite{babu2021xls} contains 317M parameters. In addition, fine-tuning can easily suffer from catastrophic forgetting and memory-based overfitting when used with datasets containing specific tasks \cite{tan2018survey}.

In recent years, the research results in the field of natural language processing (NLP) have developed rapidly, and various efficient transfer learning methods have emerged, such as adapter-based \cite{rebuffi2017learning}, prefix-based \cite{li2021prefix}, and low-rank adaptation (LoRA) \cite{hulora} techniques. These methods can achieve the most advanced effects with only partial model parameters trained, avoiding the problems brought by full-model tuning. In downstream tasks, maintaining static freezing of large-scale pre-trained language models (PLM) parameters, efficient parameter optimization methods are used to train only a small part of additional task-specific parameters, thereby alleviating extreme forgetting \cite{vander2023using} without requiring extra memory and computing resources. However, these efficient tuning methods are not systematically studied with self-supervised models in fake audio detection.

Inspired by Hu et al. \cite{hulora}, we propose a low-rank adaptive method for pre-trained models used in fake audio detection. By freezing the weights of the pre-trained model and injecting a trainable rank decomposition matrix, we reduce the number of trainable parameters to address issues during fine-tuning. We conducted extensive experiments on the ASVspoof2019 dataset, which showed that compared to global fine-tuning, our approach reduced the number of training parameters by 99.49\% with only a 0.17\% decrease in performance. Additionally, our method reduces the hardware training threshold and increases training speed.

The main contributions of this study can be summarized as follows:

\begin{itemize}
\item To our best knowledge, we are the first to apply the LoRA method to pre-trained models for fake audio detection.
\item The study found that our proposed method can significantly reduce the number of model parameters while maintaining high performance and improving training efficiency.
\end{itemize}

	\begin{figure*}[t]
	\centering
	
	\includegraphics[height=6.2cm,width=1\textwidth]{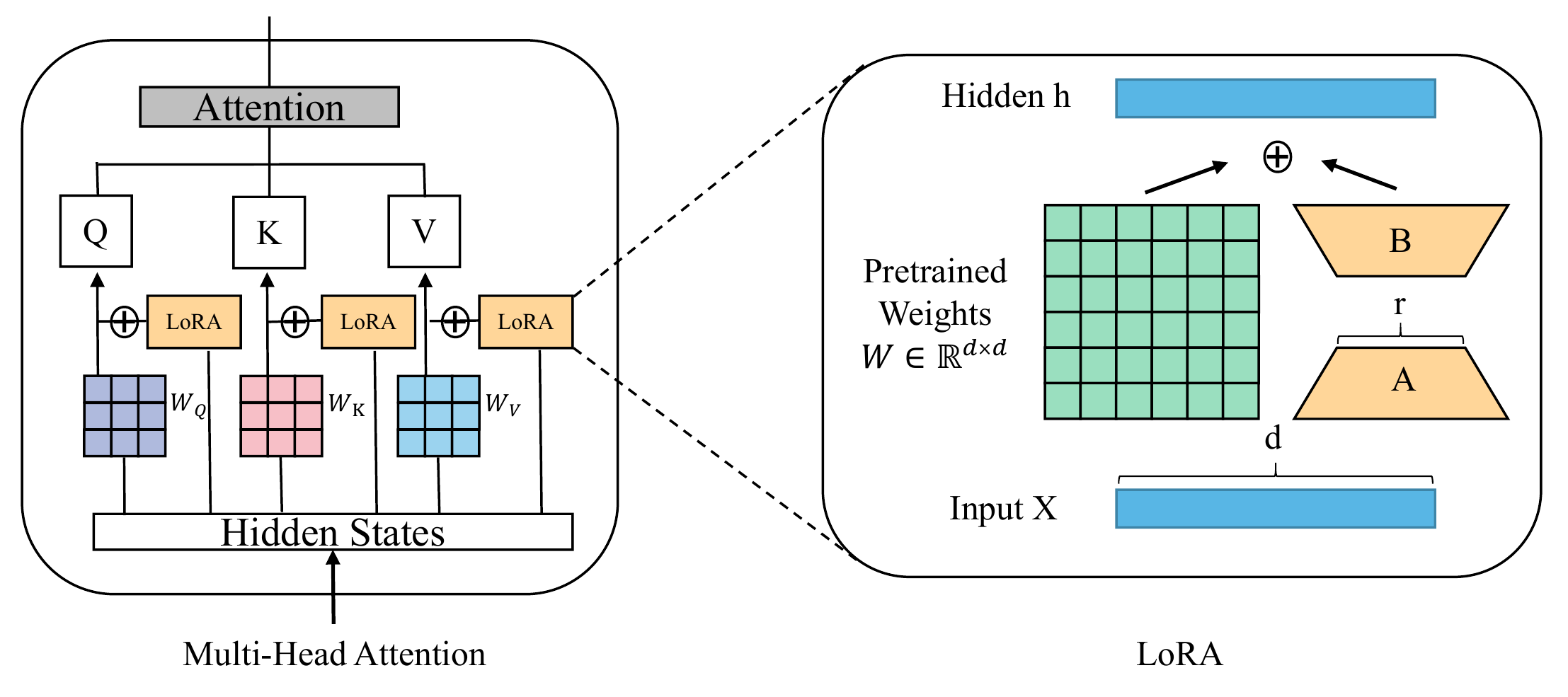}
	\caption{
		Transformer architecture in wav2vec2 along with LoRA.
	}
	
	\label{fig:speech_production}
\end{figure*}

\section{Proposed Methods}

Fine-tuning a pre-trained model is a common method used to further improve its performance on a specific task. However, the requirement that the newly fine-tuned model has the same number of parameters as the original model results in a significant increase in computing resources and higher costs. To address this issue, we were inspired by \cite{hulora} to adopt a new technique called Low-rank Adaptation (LoRA), which can reduce computational resource consumption when fine-tuning the wav2vec2 model.

To fine-tune the pre-trained weight matrix $W \in \mathbb{R}^{d \times k}$, we utilized a low-rank decomposition method to limit its update range. We express the update as $W = W + \Delta W$, where $\Delta W = BA, B \in \mathbb{R}^{d \times r}, A \in \mathbb{R}^{r \times k}$ and $r$ is the rank of the decomposition matrices with $r \ll d$. During the training process, the parameters of $W$ were fixed and did not receive any gradient updates, only the weight matrices of $A$ and $B$ were trained. It should be noted that both $W$ and $\Delta W$ have the same input vector. For the original output $h = Wx$, we obtained the refined output after low-rank decomposition as follows:

\begin{equation}
	h = h + \Delta Wx = h + BAx
\end{equation}

In this study, we applied the LoRA technique at three different locations of the multi-head attention layer in the wav2vec2 transformer, specifically at the query, key, and value vectors in the self-attention module, as shown in Figure 1. Unlike the fully fine-tuned model, LoRA introduces no inference latency and roughly converges to training the original model's performance within approximately the same number of training iterations \cite{hulora}. This means that we can significantly reduce the computational resources consumed during the fine-tuning process by applying the LoRA technique, while maintaining stable model performance.
\section{Experiments}
\subsection{Dataset}
ASVspoof 2019 LA \cite{todisco2019asvspoof} mainly has 19 spoofing attack algorithms (A01-A19), including two types of spoofing attacks: text to speech (TTS) and voice conversion (VC). The LA data set contains three subsets: the training set, the development set, and the evaluation set. Table 1 details the number of real and fake audio of the ASVspoof2019 LA dataset. The training set and development set mainly include four TTS and two VC algorithms, namely A01-A06. To better evaluate the performance of the system, unseen spoofing attacks were added to the evaluation set, including two known spoofing attacks (A16 and A19) and 11 unseen spoofing attacks (A07-A15, A17, and A18).

\begin{table}[t]
	
	\caption{ The detailed information of ASVspoof2019 LA dataset.}
	
	\centering
	\begin{tabular}{cccc }
		\toprule
		\textbf{Set} &\textbf{\#Genuine}  &  \textbf{\#Spoofed} &  \textbf{\#Total} \\
		\midrule
		Train &2,580   &  22,800  &  25,380 \\
		Dev& 2,548  & 22,296  & 24,844      \\
		Eval &7,355   & 64,578   & 71,933    \\
		
		\bottomrule
	\end{tabular}
	
\end{table}

\subsection{Evaluation Metrics}

In this work, in order to evaluate the results of different fake audio detection systems, the equal error rate (EER) is used as the evaluation metrics. Previously, EER is used in the ASVspoof challenges and ADD 2022 challenge \cite{yi2022add, todisco2019asvspoof, kinnunen2017asvspoof}. A real-valued, finite numerical value is assigned to each trial. It reflects the support for two competing hypotheses, namely that the trial is a bona fide audio or a manipulated one. But we do not optimize a decision threshold, and thus nor do we produce hard decisions. High detection score should indicate a genuine utterance and low score should indicate a manipulated utterance. The metric in this paper is the ’threshold-free’ EER, defined as follows. Let $P_{fa}(\theta)$ and $P_{miss}(\theta)$ denote the false alarm and miss rates at threshold $\theta$.

\begin{equation}
	P_{fa}(\theta) = \frac{\# \{spoof\;trials\;with\;score > \theta\}}{\# \{total\;spoof\;trials \}}
\end{equation}

\begin{equation}
	P_{miss}(\theta) = \frac{\# \{genuine\;trials\;with\;score < \theta\}}{\# \{total\;genuine\;trials \}}
\end{equation}

So $P_{fa}(\theta)$ and $P_{miss}(\theta)$ are, respectively, monotonically decreasing and increasing functions of $\theta$. The EER corresponds to the threshold $\theta_{EER}$ at which the two detection error rates are equal, $i.e. EER = P_{fa}(\theta_{EER}) = P_{miss}(\theta_{EER})$.

\begin{table}[t]
	
	\caption{Effect of the rank r of different low-rank matrices on the results of ASVspoof2019 LA eval set.}
	
	\centering
	\begin{tabular}{ccc }
		\toprule
		\textbf{r} &\textbf{\#Parameters}  &  \textbf{EER\%}  \\
		\midrule
		2 & 804k   &  2.20   \\
		\textbf{4} & 1.6M  & \textbf{1.30}       \\
		8 & 3.1M  &   1.72     \\
		16 & 6.1M  & 1.51        \\
		\bottomrule
	\end{tabular}
	
\end{table}

\begin{table}[t]
	
	\caption{Effect of the different weight matrices on the results of ASVspoof2019 LA eval set. The data in the table represents the EER\%.}
	
	\centering
	\begin{tabular}{ccccc }
		\toprule
		\textbf{Weight Type} &\textbf{r=2}  &  \textbf{r=4} &  \textbf{r=8} &  \textbf{r=16} \\
		\midrule
		$W_q$    &  2.40 & 1.54& 1.86 &1.80 \\
		$W_k$  &  2.51 & 1.82 &1.99 &1.86 \\
		$W_v$ &  2.38 & 1.47 & 1.82 & 1.75      \\
		\textbf{$W_q, W_v$} & 2.20 & \textbf{1.30} & 1.72 &1.51       \\
		$W_q, W_k$ & 2.33 & 1.45 & 1.80 & 1.68       \\
		$W_k, W_v$ & 2.33 & 1.42 & 1.82 &1.71       \\
		$W_k, W_q, W_v$ & 2.14 & 1.36 & 1.72 & 1.53       \\
		\bottomrule
	\end{tabular}
	
\end{table}

\begin{table}[t]
	
	\caption{Effect of different audio lengths on the results of ASVspoof2019 LA eval set. Train time indicate the time required for one epoch of training. To facilitate fair comparison, a standardized batch size of 8 was utilized.}
	
	\centering
	\begin{tabular}{ccc }
		\toprule
		\textbf{Length}   & \textbf{Train Time(s)}  &  \textbf{EER\%}  \\
		\midrule
		1s & 75 & 10.27   \\
		2s & 97 & 5.09   \\
		4s & 183 & 1.30       \\
		8s & 298 &  1.18     \\
		
		\bottomrule
	\end{tabular}
	
\end{table}

\begin{table}[t]
	
	\caption{The comparison of our proposed method with fixed parameters, global fine-tuning, and other local fine-tuning methods on ASVspoof2019 LA eval set. Train time indicate the time required for one epoch of training. To facilitate fair comparison, a standardized batch size of 4 was utilized.}
	
	\centering
	\begin{tabular}{cccc }
		\toprule
		\textbf{Methods} &\textbf{Train Time(s)}&\textbf{\#Parameters}  &  \textbf{EER\%}  \\
		\midrule
		Fixed & 185 & 0   &  2.56   \\
		finetune & 434 & 317M  & 1.13       \\
		Adapter\footnote{Adapter \cite{rebuffi2017learning} is a method in the field of Natural Language Processing (NLP) that reduces the number of trainable parameters during the fine-tuning process. This paper reproduces the adapter method in ASVspoof.} & 326 & 13.4M  &   1.86     \\
		\textbf{LoRA(ours)} & 202 & \textbf{1.6M}  & \textbf{1.30}        \\
		\bottomrule
	\end{tabular}
	
\end{table}

\subsection{Experimental Setup}
The wav2vec2 pretrained model variant “wav2vec2 XLSR”, which we use as a pretrained feature extractor using additional linear transformations and a larger context network, is trained on 56k hours of audio samples in 53 languages. \cite{DBLP:journals/corr/abs-2006-13979}. We chose multilingual because the paper\cite{wang2021investigating} presented that “a good self-supervised front end should be trained with diverse speech data.” The model's downsampling factor is 320. Thus, there is a 1024-dimensional vector for every 10 ms of speech. 

For the backend classifier, we chose light convolution neural network (LCNN), which is the baseline system for ASVspoof2019 and 2021. Other settings refer to the paper \cite{lavrentyeva2019stc}.

To train the model, we use the Adam optimizer with a learning rate of $1 \times 10^{-5}$. Due to the limitation of GPU memory, we set the batch size to 16. The model is trained for 50 epochs. The training set is used to train the model, the development set is used to select the model with the best performance, and finally, the evaluation set is used for evaluation.

\begin{table*}[t]
	
	\caption{ Comparison of the training efficiency between micro-adjust and Lora is listed below. The data listed in the table indicate the time(senconds) required for one epoch of training. In this experiment, we used a server equipped with NVIDIA RTX 2080Ti (11GB - 6CPU+GPU). '--' refers to out of memory.  }
	
	\centering
	\begin{tabular}{c|cccccccccc }
		\toprule
		\multirow{2}{*}{Length}& \multicolumn{2}{c}{\textbf{Batch=2}}  & \multicolumn{2}{c}{\textbf{Batch=4}} & \multicolumn{2}{c}{\textbf{Batch=8}} & \multicolumn{2}{c}{\textbf{Batch=16}} & \multicolumn{2}{c}{\textbf{Batch=32}}\\
		\cline{2-11}
		& finetune & LoRA & finetune & LoRA & finetune & LoRA & finetune & LoRA  & finetune & LoRA \\
		\midrule
		1s & 485    & 296 &248 &153 & 127& 75 &97 &46& -- & 42  \\
		2s & 416    & 303&263 &152 & 192&97&-- & 83&-- & 75   \\
		4s & 560    & 299&434 &202 & --&183&-- &170&-- & --       \\
		8s & 936    & 426&-- &375 &-- &358&-- &--&-- & --      \\
		
		\bottomrule
	\end{tabular}
	
\end{table*}

\section{Results and Discussion}

We first focus on the effect of rank r on model performance. We found that the number of model parameters is positively correlated with the value of R, indicating that the model's parameter count increases with increasing r. Surprisingly, the model achieved the best EER of 1.30 when R was set to 4, outperforming the results of the other three ranks. It is worth noting that even with a very low value of r, r=2, the model's EER was still better than when the model's parameters were fixed, demonstrating the effectiveness of combining LoRA with Wav2vec2. However, when the rank of LoRA was set to 8 and 16, the EER results were slightly worse than the rank of 4. This may be due to injecting a rank that was too large, leading to an increase in the number of model parameters and an increase in overfitting.

Table 3 shows the effect of different weight types on the results. We found that, for all weight types and low-rank matrix ranks, the LoRA model outperforms the baseline model in terms of EER, indicating that LoRA's low-rank matrix adaptation technique can effectively improve the performance of ASVspoof detection. Furthermore, for each weight type, as the low-rank matrix rank r increases, the performance of the LoRA model slightly improves, but the improvement gradually decreases. This suggests that the performance of LoRA is limited by the low-rank matrix rank, and therefore using a higher rank does not significantly improve performance. Finally, among each weight type, Wq,wv and Wq,wk,wv perform the best, indicating that applying LoRA to the query and value matrices of the Transformer can lead to better performance.

Table 4 presents the impact of input audio length on model performance. As expected, the ASVspoof detection system exhibits a continuous decrease in EER performance with an increase in audio length. Longer speech signals provide more information, facilitating the differentiation of genuine from spoofed speech. The poorest EER performance is observed at the shortest input length of 1 second, with a score of 10.27\%. This highlights the challenge of the ASVspoof detection task in short audio scenes. In contrast, the EER performance of the ASVspoof detection system drops to 1.18\% when the input length increases to 8 seconds, approaching the optimal performance. Combining the findings in Table 6, we conclude that increasing the audio length can improve the performance of the ASVspoof detection system, but it also leads to higher memory requirements during training. Therefore, it is crucial to strike a balance between model performance and training efficiency.

Based on the aforementioned experimental results, we set the rank r to 2, the length of the input audio to 4 seconds, and apply the LoRA method to the $W_q$ and $W_v$ weight matrices. Adapter \cite{rebuffi2017learning} is a method in the field of Natural Language Processing (NLP) that reduces the number of trainable parameters during the fine-tuning process. This paper reproduces the adapter method in ASVspoof. Table 5 presents the comparison of our proposed method with fixed parameters, global fine-tuning, and other local fine-tuning methods. Based on the data in the table, the following conclusions can be drawn. Firstly, both global and local fine-tuning can improve the EER performance of ASVspoof detection task compared to the method with fixed pre-trained model parameters, indicating the benefits of task-specific fine-tuning. Secondly, compared with global and local fine-tuning methods, our proposed LoRA method significantly reduces the number of trainable parameters while achieving performance comparable to global fine-tuning on the ASVspoof detection task. This indicates that the LoRA method can maintain high performance while reducing the parameter count. Finally, compared with global fine-tuning and LoRA, the performance of the Adapter method is slightly lower. This may be because Adapter only updates the weights of specific layers without updating the weights of other layers, which limits its ability to fully leverage the advantages of the pre-trained model, and therefore, performs worse than global fine-tuning and LoRA on the ASVspoof detection task. Therefore, we can conclude that the LoRA method can significantly reduce the number of model parameters by freezing the pre-trained model's parameters and injecting trainable low-rank decomposition matrices, while maintaining high performance and improving training efficiency on the ASVspoof detection task.

Table 6 presents the results of our proposed method on improving training efficiency, showing that LoRA outperforms global fine-tuning in terms of training time and memory efficiency. Specifically, for all audio lengths, LoRA's training time is significantly shorter than global fine-tuning, and this improvement becomes more pronounced with increasing batch sizes. For instance, when the audio length is 1 second, LoRA's training time only takes 46 seconds with a batch size of 16 samples, while global fine-tuning requires 1 minute and 37 seconds. Additionally, compared to global fine-tuning, LoRA can handle larger batch sizes without running out of memory. For example, when the audio length is 4 seconds, LoRA can handle a maximum batch size of 8, while global fine-tuning runs out of memory with a batch size of 4. Overall, these results suggest that LoRA is a promising approach that can improve the training efficiency of the wav2vec2 model and lower the training threshold of hardware.

\section{Conclusion}

In conclusion, the paper proposes a novel low-rank adaptation method (LoRA) to improve the efficiency and performance of the wav2vec2 model for the fake audio detection task. The experimental results show that both global and local fine-tuning can improve the performance compared to the fixed pre-trained model parameters. However, the proposed LoRA method significantly reduces the number of trainable parameters while achieving performance comparable to global fine-tuning, indicating that LoRA can maintain high performance while reducing the parameter count. Additionally, the paper finds that the performance of LoRA is limited by the low-rank matrix rank, and therefore using a higher rank does not significantly improve performance. Among each weight type, applying LoRA to the query and value matrices of the Transformer can lead to better performance. Furthermore, increasing the audio length can improve the performance of the ASVspoof detection system, but it also leads to higher memory requirements during training. Finally, LoRA outperforms global fine-tuning in terms of training time and memory efficiency, making it a promising approach to improve the efficiency and performance of the wav2vec2 model for the fake audio detection task.

\bibliography{sample-ceur}

\appendix

\end{document}